# EL MOVIMIENTO DE LOS CIELOS

## Una propuesta pedagógica para docentes de la escuela secundaria

**Alejandro Gangui**
Instituto de Astronomía y Física del Espacio, UBA-CONICET

*Sumario para* **Contenidos**

Secuencia didáctica sobre temas de astronomía destinada a docentes de la escuela secundaria.

Con esta nota, CIENCIA HOY vuelve a la práctica de proporcionar a docentes del secundario sugerencias y orientaciones sobre cómo presentar en el aula determinados temas científicos, o cómo aprovechar mejor para su labor pedagógica lo publicado en la revista.

Lo que sigue presenta una secuencia didáctica acerca de algunas cuestiones de astronomía. Está destinada a docentes de la escuela secundaria, para que la lleven al aula. Su tema central, a ser presentado en forma progresiva y secuenciada, es *el movimiento de los cielos* o –si se prefiere enfatizar otro aspecto– *los movimientos de la Tierra*. Entre las actividades que se sugieren se cuenta que los alumnos construyan un modelo o maqueta simple de cómo se mueve nuestro planeta con relación al Sol. Ello los ayudaría a compenetrarse con aspectos astronómicos a veces tratados muy por encima en la escuela o sencillamente omitidos, por falta de recursos de fácil empleo.

El propósito del trabajo no es difundir los esquemas teóricos de pensamiento utilizados por los científicos en la actualidad, sino aprovechar la diversidad de concepciones que puedan surgir con el desenvolvimiento de las actividades y seleccionar las que mejor respondan a las preguntas que se planteen los mismos alumnos o les sugiera el docente. Es en este proceso de discusión y reflexión que se converge naturalmente a los modelos que hoy acepta la ciencia, los que, como bien sabemos, no son permanentes sino que se hallan sujetos a un continuo y saludable proceso de refinamiento y revisión.

Este trabajo se refiere, en principio, al eje 'En relación con la Tierra, el Universo y sus cambios' de los núcleos de aprendizajes prioritarios de ciencias naturales prescriptos para primero y segundo años, si bien ello es relativo porque los diseños curriculares de muchas jurisdicciones están actualmente en revisión, y porque nada impide aprovechar la propuesta en años superiores. En la actualidad la astronomía y cosmografía están poco representadas en la educación secundaria argentina; la responsabilidad de su dictado recae en profesores de física, matemática o geografía, o en aquellos docentes de buena voluntad con ganas de abordarlas como proyectos especiales. El autor se congratula de conocer a varios de esta última clase.

Las tres actividades que se proponen en lo que sigue difícilmente ofrezcan dificultades insalvables a docentes de secundaria, aunque el modo de trabajo con los alumnos dependerá del tiempo y el esfuerzo que cada uno esté dispuesto a dedicarles, y de los intereses y aptitudes del grupo particular de estudiantes de que se trate, que solo el docente es capaz de conocer.

Hemos optado por abstenernos de dar una indicación del tiempo que ocuparían las actividades que siguen, en módulos, horas o jornadas. El requerido para cada una dependerá de la capacidad de abstracción y del conocimiento geométrico y astronómico de cada grupo de estudiantes, distintos entre quienes comienzan y quienes terminan la secundaria. El docente, además, sabe bien que los tiempos dependen de la dinámica de cada grupo de alumnos, del grado de profundidad conceptual que procure alcanzar en el tratamiento de cada tema y de sus objetivos didácticos.

En un número próximo de CIENCIA HOY presentaremos otras tres actividades, de mayor complejidad, que serán continuación natural de las delineadas aquí pero apuntarán a alumnos de los últimos años con alguna predisposición, y quizás también cierto entrenamiento, para el pensamiento abstracto y reflexivo, y para la construcción de modelos. En esas futuras actividades trabajaremos con la forma real de la Tierra, cuya rotación la apartó de una esfera perfecta, y con un sutil bamboleo que posee su eje de rotación, que los astrónomos hacen responsable de lo que dieron en llamar la *precesión de los equinoccios*.

Es interés del autor que los docentes curiosos e interesados por la astronomía puedan, por lo menos, aprovechar una parte de sus propuestas, independientemente del año en que enseñen, del nivel socio-económico de la escuela y de los restantes temas tratados en esta u otras áreas. Mientras se observa, se discute, se reflexiona y se lleva a cabo una construcción simple, la secuencia de actividades sugerida no es más que una excusa para hablar sobre astronomía.

## Material de trabajo

Para llevar adelante esta propuesta se requiere disponer del siguiente material:

* Dos pequeñas esferas de telgopor, de unos 6 y 12cm de diámetro respectivamente.
* Dos esferas más grandes del mismo material, que dupliquen el tamaño de las anteriores.
* Una base cuadrada de cartón rígido, de unos 30cm de lado y algunos milímetros de espesor.
* Dos varillas delgadas de madera, de sección circular, de unos 8 y 16cm de largo respectivamente, con sus extremos puntiagudos para poder clavarlas al telgopor.
* Un poco de plastilina, un delgado tubo de plástico de unos 5cm de largo (serviría cortar un tanque vacío de bolígrafo) y unos 40cm de hilo de algodón de envolver.
* Dos alambres rígidos y rectos, de aproximadamente 15cm de largo.
* Tijera, cinta engomada y marcador de color oscuro.
* Una arandela y un pequeño cubo de telgopor de 2cm de lado.

## Secuencia de actividades

### Actividad 1: El ciclo día-noche, el año y los movimientos de la Tierra

El objetivo de esta actividad es iniciar a los alumnos en una reflexión acerca de la rotación de la Tierra sobre su eje y su traslación anual en órbita alrededor del Sol. Búsquese que expresen sus ideas y discutan en grupo. Ante un conflicto entre esas ideas y lo que resulte de la reflexión y la discusión grupal, procúrese que tiendan a converger por sí mismos hacia los conceptos que hoy maneja la ciencia.

Se inicia la actividad empleando las dos esferas mayores de telgopor (las dos más pequeñas quedan para la última actividad), para representar con ellas, respectivamente, el Sol y la Tierra (la segunda, como es obvio, por la esfera de 12cm de diámetro). Se comienza con el ciclo día-noche, vale decir, explorando en forma grupal qué ocasiona la sucesión de los días y noches sobre la superficie de nuestro planeta.

Un pequeño número de preguntas planteadas por el docente puede ayudar a guiar la discusión del grupo. Preguntas posibles son:

- * ¿Por qué hay días y noches? ¿Qué determina el ciclo día-noche?
- * Este ciclo, ¿es el mismo para todo habitante de la Tierra?
- * ¿Existen regiones donde, a veces, pasan muchos 'días' sin luz directa del Sol? ¿Existe el *país de las sombras largas*? ¿Existen regiones donde, a veces, el Sol no se pone de 'noche'?
- * De noche, los habitantes de la Tierra:
  - ¿Ven las mismas estrellas que verán, 12 horas más tarde, sus *antimeridianos*? (Antimeridianos son quienes viven en localidades con igual latitud pero con 180° de diferencia en longitud: el antimeridiano de Buenos Aires está en el Océano Índico, apenas al este de West Cape Howe, en Australia.)
  - ¿Ven las mismas estrellas que verán, 12 horas más tarde, sus *antípodas*? (Literalmente antípoda significa *opuesto por los pies*. Igual que con los antimeridianos, con los antípodas hay una diferencia en longitud de 180° y, además, en casi todos los casos, una en latitud; los antípodas tiene la misma latitud pero están en el otro hemisferio: los antípodas de los porteños (habitantes de la ciudad de Buenos Aires) nadan en el Mar Amarillo, hacia el oeste de Corea del Sur. ¿Entre qué antípodas no hay diferencia de latitud y solo de longitud?).

Las preguntas son solo una guía para provocar la discusión grupal y para producir con las esferas de telgopor una suerte de sencilla dramatización de los fenómenos que se analizan. Aunque por lo general se considera que el tema de esta actividad es simple (y hasta obvio), hay dos razones para no restarle importancia. En primer lugar, permite que el docente conozca algunas de las *ideas previas* o *preconceptos* (también llamadas *ideas naturales*) de sus alumnos. Las cabezas de los alumnos no están vacías, por más que haya quienes, con cariño o desconsideración, los llamen 'cabezas huecas'. Como todos nosotros, ellos perciben el mundo que los rodea e inventan modelos mentales más o menos correctos de su funcionamiento. Si sus ideas previas no coinciden con las que acepta la ciencia y mejor explican las observaciones, constituyen un obstáculo para el buen aprendizaje. Conocer lo que piensan los alumnos sobre un fenómeno, ya sea astronómico o de cualquier otra índole, ayuda al docente a encarar la enseñanza sobre una base mejor.

En segundo lugar, cuando se abordan temas de este tipo, que rápidamente se vuelven más abstractos y, por ende, dejan de ser obvios (como veremos), siempre es mejor comenzar con lo simple y obvio.

La discusión en grupos pequeños y el intercambio de ideas entre pares, además de evidenciar qué piensan los alumnos sobre el tema, conduce, con la guía del docente, a una representación más fiel de los fenómenos que se intentan explicar.

En un momento apropiado, y si los estudiantes no lo han mencionado, el docente les podrá sugerir que identifiquen ciertas partes distintivas de la superficie terrestre, especialmente los polos, por los que pasa el eje imaginario de rotación de la Tierra. Con una varilla de madera larga que atraviese diametralmente la esfera de telgopor de la Tierra se materializa ese eje de rotación. Ello ayuda a los alumnos a reflexionar sobre el movimiento rotativo de la Tierra y a imaginar mejor su efecto sobre el ciclo día-noche.

Aquellos alumnos que, cuando intentan explicar los días y las noches, confunden la traslación de la Tierra alrededor del Sol con la rotación del planeta sobre su eje, por lo común advierten al serles planteadas estas u otras preguntas similares la inconsistencia o conflicto conceptual que se les presenta, y se enfrentan con la necesidad de resolverlo. Por ejemplo, si suponen que la Tierra no rota pero se traslada alrededor del Sol y cumple una vuelta completa del astro en 24 horas, dan con un modelo que explicaría en principio el ciclo día-noche. Sin embargo, ese modelo, entre otras cosas, no permite aclarar cómo suceden los largos pero poco luminosos días o las heladas e interminables noches que, separadas por unos seis meses, se alternan en las regiones polares (para la explicación de estos fenómenos se requiere tener en cuenta el ángulo del eje de la Tierra con relación a su plano de traslación, un concepto al que se dedica la segunda actividad).

Luego de las discusiones preliminares, en que se habla esencialmente sobre el movimiento de rotación de la Tierra, el docente puede concentrar la atención de los estudiantes en el movimiento anual de traslación del planeta en torno al Sol, y alentarlos a que lo exploren. Es posible considerar diferentes alternativas: que la trayectoria de dicho movimiento carezca de forma definida, que esté contenida en un plano, que la órbita descripta sea una circunferencia, que sea una circunferencia achatada o *elipse* y, si fuera lo segundo, con qué grado de chatura o *excentricidad*.

Algunas posibles preguntas para guiar la discusión serían:

* Considerando solo a la Tierra y al Sol, ¿qué cuerpo se mueve alrededor del otro?
* ¿Qué elementos tienen para justificar esta respuesta? Expliquen.
* ¿Cómo creen que es el movimiento anual de la Tierra alrededor del Sol?

La respuesta a la primera pregunta depende del sistema de referencia implícito o explícito en la descripción del fenómeno. Así, si solo se intenta explicar la sucesión de los días y las noches y se prescinde de todo lo demás que hay en el cielo, definir un sistema caracterizado por una Tierra que no rote sobre su eje y un Sol que la circunvale (y cumpla un giro cada 24 horas) proporciona una explicación razonable. Pero si incluimos en nuestro modelo mental no solo a la Tierra y al Sol sino, también, a los demás planetas del sistema solar, la combinación de rotación de la Tierra en torno a su eje con su circunvalación del Sol proporciona una respuesta mucho más satisfactoria a esa primera pregunta. Recuérdese que nunca hubo una forma sencilla de definir las trayectorias de los planetas sobre la base de considerar que ellos y el Sol se mueven alrededor de la Tierra (es decir, de construir un modelo geométrico de su órbitas vistas desde la Tierra).

Luego de que los alumnos den sus respuestas a la segunda pregunta (las que quizás podría el docente registrar en el pizarrón), y antes de pasar a la tercera pregunta, se puede sugerir la lectura del recuadro 'Una anécdota sarmientina' (ver más abajo), y hasta organizar un debate entre alumnos que representen a los personajes de ese texto. Siendo una anécdota, es muy posible que la trascripción no haya sido demasiado fiel: donde dice 6 ½ leguas *por minuto* debería decir *por segundo*, valor sesenta veces mayor que, en realidad, es cercano a los aproximadamente 30km/s que podemos hoy determinar cuando adoptamos el modelo de la Tierra desplazándose en órbita alrededor del Sol y calculamos la velocidad de ese desplazamiento.

Como esta primera actividad aborda conceptos que quizás los alumnos ya dominen o hayan adquirido hasta cierto punto en clases anteriores, el docente podrá establecer si cree necesario plantear todas las preguntas y entrar en todas las cuestiones enumeradas. Pero antes de pasar a la siguiente actividad conviene terminarla reuniendo a todo el grupo y haciendo una puesta en común para terminar de fijar y organizar los conocimientos.

## Actividad 2: Las cuatro estaciones

El objetivo de la segunda actividad es que los alumnos reflexionen sobre la orientación en el espacio del eje de la Tierra, y reconozcan que una Tierra cuyo eje sea perpendicular al plano de su órbita no permite explicar la existencia de las estaciones del año. Se aconseja trabajar en grupos pequeños a partir de preguntas-guía planteadas por el docente.

En esta actividad se trabaja también con las esferas grandes que representan al Sol y la Tierra, y se procura asimismo promover la discusión entre los alumnos y hacer aflorar sus ideas previas sobre los fenómenos en discusión. El docente puede sugerir que en cada grupo los estudiantes desplacen la Tierra a lo largo de su órbita y alentarlos a que consideren orientaciones alternativas del eje terrestre. Con ello habrá diferencias de opinión en el seno de cada grupo. Ciertas preguntas planteadas por el docente pueden resultar útiles para guiar la discusión, por ejemplo:

* ¿Qué determina las estaciones del año?
* ¿Porqué cada estación no ocurre en el hemisferio norte en la misma época del año que en el hemisferio sur? Justifiquen sus respuestas.
* ¿Qué debería sucederle al eje de la Tierra para que el Sol se elevara por encima del horizonte alguna vez en el año en los polos?

En la mente de mucha gente, incluidos estudiantes, la causa de las estaciones es la mayor o menor distancia, en diferentes momentos del año, entre la Tierra y el Sol, nuestra fuente de luz y calor. Esta noción, que se suele denominar la teoría del alejamiento, postula que en invierno la radiación solar llega desde más lejos que en verano. Varios factores explican tal concepción. Es cierto que hay momentos del año en que la Tierra y el Sol se acercan (el punto de la órbita terrestre llamado *perihelio* es el de máxima cercanía) y otros en que se alejan (el *afelio* es el punto de máxima lejanía), pero, aparte de que las diferencias de radiación recibida no son significativas, ello no puede explicar por qué cuando en un hemisferio terrestre es pleno verano en el otro es, simultáneamente, pleno invierno.

Posiblemente la razón más importante de este error sea la costumbre de dibujar en perspectiva, en libros y revistas, la órbita de la Tierra, lo que exagera su forma elíptica y magnifica al extremo los cambios de la distancia al Sol a lo largo del año. Es importante que el docente esté prevenido de lo difundida que es esta noción, y que aproveche la discusión de las preguntas sugeridas para que los estudiantes arriben a una explicación más satisfactoria de las estaciones del año.

Con la guía del profesor, esa explicación se centrará en el concepto de la inclinación del eje de rotación de la Tierra, que no es perpendicular al plano de la eclíptica: forma con esa perpendicular un ángulo de 23°26' (23,5 grados aproximadamente). El docente puede aclarar en el momento oportuno, de paso, que el plano definido por la órbita de la Tierra en su viaje alrededor del Sol se denomina *eclíptica* (pues solo pueden producirse eclipses cuando la Luna cruza ese plano en su camino).

Desde que se enfrentan con las primeras preguntas acerca de esta actividad, los alumnos probablemente encuentren útil marcar en la superficie de la representación de la Tierra, como guía, ciertos elementos característicos, como el ecuador, los trópicos o los círculos polares, que pueden dibujar con el marcador. También, materializar el eje de rotación de la Tierra mediante una varilla clavada de polo a polo los ayudará a razonar.

## Actividad 3: Construcción de un mini sistema solar

Con esta actividad se busca que los alumnos apliquen los conceptos de la inclinación del eje terrestre y de la órbita casi circular de nuestro planeta a construir una maqueta simple que ilustre los movimientos relativos de la Tierra y el Sol. El dispositivo solo incluye al Sol y a la Tierra, con esta como satélite del astro.

Igual que las anteriores, la tercera actividad se realiza con los alumnos distribuidos en grupos, cada uno de los cuales construye un modelo o maqueta con los materiales indicados al comienzo de la nota, del modo que indica la figura 1. El profesor recorre los grupos, primero para dar indicaciones sobre cómo construir el dispositivo, y luego para discutir detalles de la maqueta. Les podrá hacer notar, entonces que, en el modelo, la órbita de la Tierra es una circunferencia, y no una elipse como en la realidad.

Será también una buena oportunidad para aclarar que la órbita terrestre real no se aparta mucho de una circunferencia, y que ese apartamiento se mide por un valor llamado *excentricidad*. La excentricidad de una circunferencia es cero; una elipse achatada al extremo, que aparecería como dos segmentos rectos que tienden a confundirse, tendría una excentricidad enormemente cercana a 1. La excentricidad de la órbita terrestre es tan solo 0,0167. Por ello, no se deforma en exceso la realidad si, para fines didácticos, se usa en el modelo una órbita circular.

Otro detalle de interés para poner de relieve es que, en su movimiento anual alrededor del Sol, la orientación del eje terrestre no se modifica; es decir, los extremos de dicho eje siempre apuntan en las mismas direcciones del espacio, o hacia los mismos puntos de la esfera celeste. En el hemisferio norte, ese punto resulta materializado por la estrella *Polaris* o polar, una estrella brillante que no se puede ver desde la Argentina. En el hemisferio sur no hay una estrella polar brillante, y el punto virtual se ubica aproximadamente prolongando imaginariamente en el cielo uno de los brazos de la Cruz del Sur. En clase, ese punto o estrella polar norte puede ser reemplazado por una mancha u objeto convenientemente ubicado en el techo (algo desplazado de la vertical de donde se instale la maqueta).

Con lo avanzado hasta aquí habrá bases suficientes para comprender qué son los trópicos y los círculos polares. Los alumnos estarán en condiciones de apreciar que existe en la Tierra una banda que podríamos llamar *tropical*, a ambos lados del ecuador, entre las latitudes de 23º26' norte y 23º26' sur, donde el Sol se ubica a mediodía en el cenit en algún momento del año. Las mencionadas latitudes son, respectivamente, las de los trópicos de Cáncer y de Capricornio. El profesor podrá preguntar cuándo y dónde se vería el Sol de mediodía en el cenit si el eje de rotación de la Tierra no estuviera inclinado sino perpendicular a la eclíptica. La respuesta completa no solo tendría que explicar que ello sucedería sobre el ecuador sino, también, todos los días del año.

Se pueden usar las esferas grandes de telgopor para marcar los trópicos y los círculos polares ártico y antártico (estos situados, respectivamente, en las latitudes 66º34' norte y sur, es decir, a 23º26' de los polos), cuyas ubicaciones quedan determinadas por la inclinación del eje de rotación de la Tierra. Con adecuadas preguntas el docente llevaría a comprender que fuera de la banda tropical, es decir, en latitudes mayores que la de los trópicos, nunca pasa el Sol de mediodía por el cenit; y que en las bandas polares, por encima de las latitudes de los círculos polares (y solo allí), se producen los fenómenos del sol de medianoche y de días en que no sale el sol. Esta simple construcción permite discutir nociones de uso cotidiano relacionadas con las estaciones, como los solsticios y equinoccios, y explicar con la maqueta su verdadera significación astronómica.

De manera similar a lo que se sugirió antes, conviene terminar esta actividad reuniendo a todo el grupo para redondear los conocimientos adquiridos y quedar en condiciones de dar el siguiente paso que explicaremos en un número venidero [ver GANGUI A, 2008, 'La precesión de los equinoccios', *Ciencia Hoy*, 18, 107:54-63.]

*Recuadro*

### Una anécdota sarmientina

En 1842, Domingo Faustino Sarmiento fue designado primer director de la Escuela Normal para Maestros de Chile. Enseñaba allí, además, los que denominaba 'cursos científicos'. En una oportunidad, un alumno de una clase de cosmografía en que explicaba la estructura heliocéntrica del sistema planetario interrumpió la explicación para objetar sus fundamentos: *Yo no creo lo que usted está diciendo. Cuando más lo admitiré como una hipótesis* afirmó. *Muy bien*, respondió el profesor. *¿Sabe usted cuánta distancia hay de la Tierra al Sol?* El estudiante contestó que conocía esa distancia. *¿Y de la Tierra a las estrellas?* insistió Sarmiento. Acordaron que es inmensa. El profesor argumentó entonces: *Fije usted una cifra en millones de leguas. Si la Tierra no gira en torno del Sol, las estrellas giran en veinticuatro horas en torno de la Tierra. Esa distancia es el semidiámetro de un círculo; luego, multiplicando el semidiámetro por seis, obtendré aproximadamente el espacio que usted hace recorrer a las estrellas por día, por hora, o por minuto; es decir, muchos millones de leguas por minuto; mientras la teoría contraria le da 6 ½ leguas por minuto de marcha de la Tierra en torno del Sol, lo que hace una parte de la distancia que recorren los ferrocarriles. Así, pues, la verdad es verosímil, mientras que el sistema de usted es absurdo e inútil. ¿A qué fin han de dar esta inconcebible vuelta, y hasta lo infinito de velocidad en los extremos, en torno de nuestro globo todos los días*? (A Belín Sarmiento, *Sarmiento anecdótico*, Saint Cloud, 1929, citado por Marcelo Montserrat, 'Sarmiento y los fundamentos de su política científica', en Miguel de Asúa (ed.), *La ciencia en la Argentina, perspectivas históricas*, CEAL, Buenos Aires, 1993).

## *Datos del autor*

Alejandro Gangui

Doctor en astrofísica,
International School for Advanced Studies, Trieste, Italia.
Investigador adjunto del CONICET. Profesor de la FCEyN, UBA.
Miembro del Centro de Formación e Investigación en Enseñanza de las Ciencias, FCEyN-UBA.
*gangui@df.uba.ar*
*cms.iafe.uba.ar/gangui*

*Figuras*

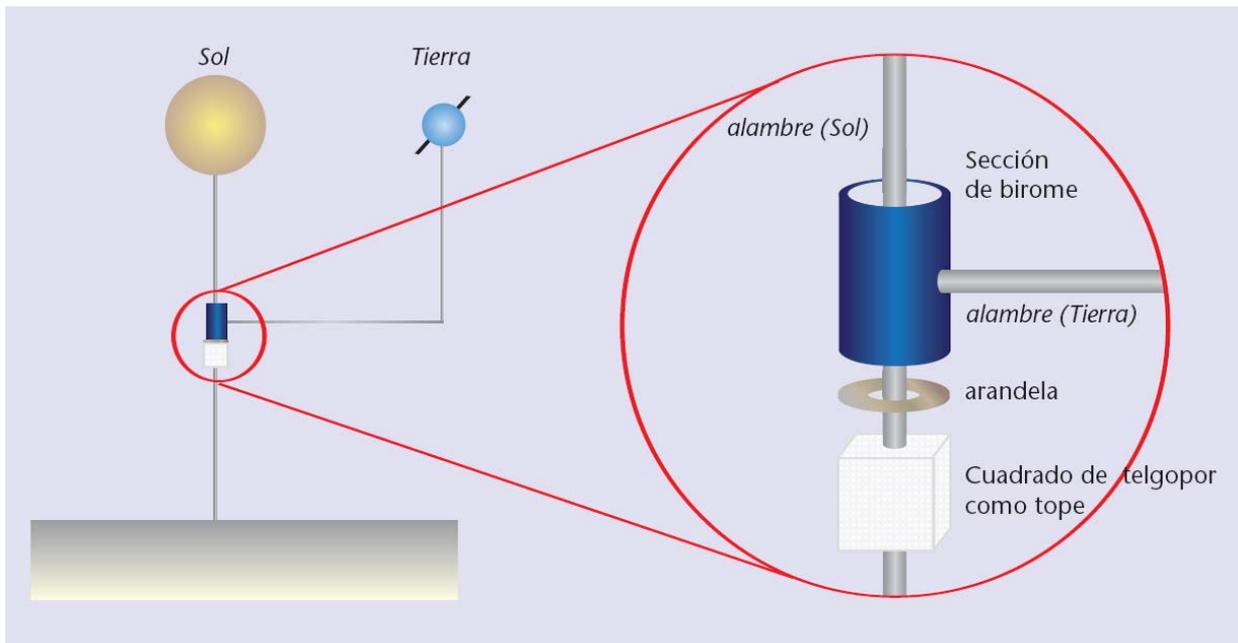

**Figura 1**. Maqueta de un sistema solar simplificado compuesto por el Sol y la Tierra, construida sin tomar en cuenta la escala con los materiales indicados en el texto. Sobre la base de cartón, empleando un trozo de plastilina, se fija verticalmente uno de los alambres rectos. En su extremo superior se inserta la esfera de telgopor de 12cm de diámetro, que hace las veces de Sol. Se dobla otro alambre en ángulo recto y se fija un pequeño tubo de plástico (por ejemplo, un trozo de tanque de bolígrafo) en uno de sus extremos. En el otro extremo se inserta la esfera de telgopor de 6cm de diámetro, que representa a la Tierra, atravesada por la varilla de madera de 8cm que representa su eje de rotación. El croquis de detalle (derecha) muestra la unión del alambre que sostiene a la Tierra con el tubo de plástico, el cual permite que la Tierra se traslade alrededor del alambre que sostiene al Sol. Por debajo de dicho tubo y sin sujetarla a él se coloca una arandela sostenida por el cubo de telgopor fijado con plastilina al alambre vertical, de suerte que evite que el dispositivo que sostiene a la Tierra deslice hacia abajo pero permita que rote alrededor del Sol. Además pueden fijarse a la plastilina que sostiene el cubo de telgopor y al cartón de la base, a modo de vientos de una carpa, cuatro hilos de algodón que den firmeza a la maqueta.

Adviértase que la esfera de telgopor de la Tierra no puede estar pegada al alambre que la sostiene, sino que debe poder rotar con relación a este para que se mantenga constante la orientación del eje de la rotación real del planeta. En la maqueta, la esfera de telgopor rotará sobre el alambre que la sostiene una vuelta completa por cada revolución anual que efectúe la Tierra, un movimiento inexistente en la realidad porque allí ningún alambre sostiene al planeta. En la maqueta, ese alambre impide que se simule la rotación diaria de la Tierra sobre su eje.

Adviértase también que el lugar de inserción del alambre en la esfera de la Tierra deberá ser el apropiado para asegurar que esta quede con su eje inclinado con respecto a la base de cartón (en esta configuración, el plano del cartón es paralelo al de la eclíptica). De esta manera, el eje inclinado de la Tierra permitirá explicar las estaciones del año, mientras que el movimiento de rotación del alambre de la Tierra alrededor del alambre vertical del Sol permitirá simular la órbita anual terrestre.

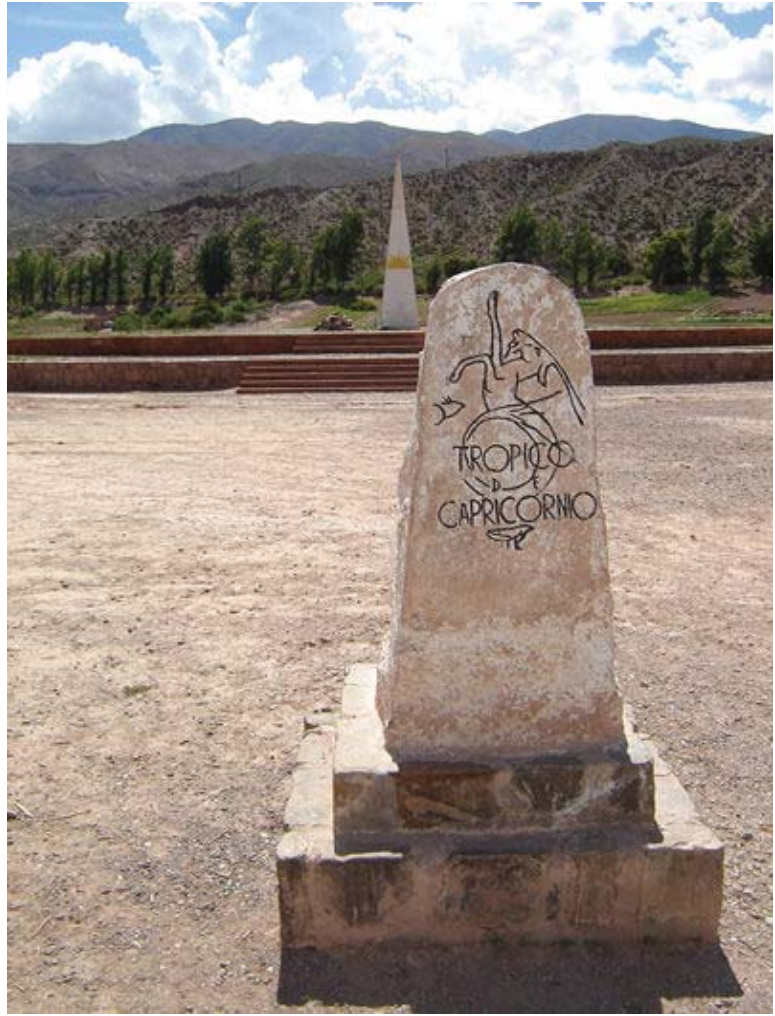

**Figura 2**. Monumentos que marcan la latitud del trópico de Capricornio en las cercanías de Huacalera, en la quebrada de Humahuaca, Jujuy.

*Lecturas sugeridas*